\documentstyle[12pt,epsfig]{article}
\begin{document}
\begin{flushright}
LPT-ORSAY 01/07 \\
NYU-TH 01/01/02 \\
hep-th/0101234
\end{flushright}
\vskip 1cm
\begin{center}
{\Large \bf The radion in brane cosmology}\\
\vskip 2cm
{Pierre Bin\'etruy$^*$, C\'edric Deffayet$^\dagger$, 
David Langlois$^{\dagger \dagger}$\footnote{
Pierre.Binetruy@th.u-psud.fr, cjd2@physics.nyu.edu, langlois@iap.fr}}\\
\vskip 1cm
$^*$ LPT\footnote{Unit\'e mixte de recherche UMR n$^o$ 8627.},
Universit\'e Paris-XI, B\^atiment 210,
F-91405 Orsay Cedex, France;\\
$^\dagger$ Department of Physics, New York University,\\
4 Washington Place, New York, NY 10003, USA.\\
$^{\dagger \dagger}$  Institut d'Astrophysique de Paris, \\
Centre National de la Recherche Scientifique,\\
98bis Boulevard Arago, 75014 Paris, France\\
\vskip 1.6cm
\end{center}

\def\beq{\begin{equation}}
\def\eeq{\end{equation}}

\def\ro{{\rho_2}}
\def\P{{P_2}}
\def\y{{\cal R}}
\def\dy{{\delta\cal R}}
\def\dS{{$dS_{4}$}}
\def\AdS{{$AdS_{4}$}}
\def\d{{\delta}}

\abstract
We consider the homogeneous  cosmological radion, which we 
define as  the interbrane distance in a
two-brane and  $Z_2$ symmetrical configuration.
In a coordinate system where one of the brane is at rest, 
  the junction conditions for the second (moving) brane 
give  directly the (non-linear) equations of motion for the radion. 
We  analyse the radion fluctuations and solve 
the  non-linear dynamics in some simple cases of interest.
%**********************************************************************
\section{Introduction}
%**********************************************************************
Motivated by recent 
developments in  string theory, the idea of confining matter 
in a three-brane embedded in a higher dimensional spacetime has stimulated
lately a lot of activity in several fields of research. 
In cosmology, it has been shown that, quite generically, the matter 
confinement in a brane entails a non-standard evolution of the brane 
geometry \cite{bdl99}. 
In other words, the Friedmann equations governing the expansion 
of our `apparent' universe are modified. 
This could lead to trouble because nucleosynthesis would give very different 
light element abundances with a modified scale factor evolution. 
Therefore, a strong constraint on brane models is  the recovery 
of  standard cosmological evolution at least since nucleosynthesis.
A nice solution to this problem has been provided by applying the idea 
of the  Randall-Sundrum model \cite{rs99a,rs99b}
  that a tension in the brane can be compensated by 
a negative cosmological constant in the bulk, as far as the internal brane 
expansion is concerned. Applied to cosmology, this idea leads to the recovery 
of the usual Friedmann law when the ordinary cosmological energy density
is much less than the brane tension \cite{cosmors}. 
By contrast, at earlier times, when 
the energy density is higher than the tension one finds a non-standard 
evolution in the brane \cite{bdel99}.

Another solution that has been suggested in the case of finite size
extra-dimension (see e.g. \cite{ced} for the case of infinite extra-dimension)
is to consider a stabilizing potential for the so-called radion
\cite{cgrt99,Kanti99}. The precise definition of the radion
is the source of some confusion in the literature. It is usually described  as
the metric component along the extra-dimension(s), but as emphasized in
\cite{cgr99} it can be reinterpreted as the relative distance between the two
branes. Stabilization of the radion can  be
explicitly realized  \cite{gw99}
by adding some matter in the bulk, a scalar field for simplicity, 
so that the relative distance between the two branes will be stabilized.
This mechanism has been applied \cite{gw99,DeWolfe:2000cp,tm00,gpt00}
 to the two-brane model of Randall and 
Sundrum \cite{rs99a}, in order to recover gravity consistent with the 
observations on the negative tension brane (without stabilization, one gets
a Brans-Dicke gravity with a negative parameter \cite{gt99}).
 In cosmology,
this mechanism has been discussed in \cite{cgk00} and \cite{cf00} but  
the claim that this  would lead to the usual 
Friedmann equations has been challenged in \cite{mb00}. Let us mention
that the radion has also been discussed in the context of holography
\cite{holography}.

The purpose of the present work will not be to study the cosmological 
consequences of a stabilizing mechanism for the radion, task which will 
be left for future work, but more modestly to characterize the radion 
in cosmology (see also a brief discussion in \cite{bcg00}), to write down the equations of motion that govern its 
dynamics, which will be analysed in some simple examples.

Previous treatments of the radion in brane systems \cite{cgr99,tm00,gs00,kr00}
start from the beginning with a perturbative approach: the radion is thus
hidden among the metric fluctuations, which have been the subject of active
research lately in cosmology (see \cite{l01} and references therein).
Our approach is quite different  in the sense that the 
radion dynamics appears directly from the junction conditions. It also gives
us the full non-linear dynamics of the homogeneous radion (i.e. ignoring 
the ordinary spatial dependence of the radion). We are thus able to show that
the dynamics of the radion is more subtle than that of a scalar field 
coupled to the trace of the matter energy-momentum tensor. 

Our plan is the following. We begin in section 2 by defining the
coordinate system
in which we describe the two-brane configuration. We then focus on the
internal geometries inside the two branes in section 3. Section 4 is the 
central part
of this paper, where the equations of motion of the radion are derived. 
Starting in section 5, we apply our general results to the case of an 
empty anti-de 
Sitter ($AdS_5$) bulk spacetime.
The last two sections are  devoted to the 
analysis  of the radion evolution in this geometry, first using a
perturbative approach (section 6), then in the full non-linear case 
(section 7). 

%**********************************************************************
\section{The two-brane cosmological system}
%**********************************************************************
The purpose of this section will be to present in a geometrical way the 
space-time configuration we wish to study. The spacetime will be  supposed to
be five-dimensional and to  
satisfy three-dimensional spatial homogeneity and isotropy in order to 
obtain (unperturbed) cosmological configurations. Two 
spatially homogeneous and isotropic three-branes live in this spacetime. The 
fifth dimension is periodic and we assume a mirror (orbifold) symmetry 
across each of the branes. 
The space between the two branes represents {\it half} of the periodic space
along the fifth dimension, the other half being the mirror image with 
respect to any of the two branes.

In any coordinate system, even  adapted to the three-dimensional 
homogeneity and isotropy, 
 the two branes would be in general in motion along the fifth dimension.
It turns out that it is possible, and convenient, to define the fifth 
dimensional  coordinate $y$ 
as part of a  Gaussian normal 
coordinate system with respect to  one of the branes, chosen 
arbitrarily and called the {\it reference brane}. The reference brane is at 
rest   in the fifth 
dimension, in $y=0$ say, and the 
 spacetime metric in this coordinate system (which is  not 
unique) is of the form 
\beq
  ds^2=g_{AB} dx^A dx^B=
-n(t,y)^2 dt^2+a(t,y)^2 \gamma_{ij}dx^i dx^j+ dy^2, \label{metric}
\eeq
where the $x^i$ are coordinates for the maximally symmetric three-dimensional 
submanifolds, $t$ is a time coordinate. 
The metric components are 
necessarily of this form because of the symmetries, but their explicit 
dependence of $t$ and $y$ can be obtained only by solving the 
five-dimensional Einstein's equations (with the appropriate boundary 
conditions),  
\beq
G_{AB}\equiv R_{AB}-{1\over 2}R g_{AB}=\kappa^2 T_{AB},
\label{einstein}
\eeq
where $R_{AB}$ is the Ricci tensor, $R$ its trace, and $T_{AB}$ 
the energy-momentum tensor. The coupling of the matter to  the 
geometry, $\kappa^2$, can be written as 
 the inverse of the third power of the five dimensional reduced
Planck mass.

If the reference brane is now at rest, the second brane 
will be in general moving in the same coordinate system. 
Its  position  at any time $t$ will be denoted 
\beq
y=\y(t),
\eeq
and we call this function $\y(t)$, which represents the relative distance 
between the two branes,  the {\it cosmological radion}. 
Note that the r\^ole 
of the reference brane and of the second brane can of course be exchanged. 
Finally we wish to stress that here, 
unlike  most of the literature, our radion is defined in a 
non-perturbative way. We will show later that our definition coincides, 
in the linearized approach, with the traditional definition of the radion.

%**********************************************************************
\section{Induced geometries in the two branes}
%**********************************************************************
The induced metric in the reference brane immediately follows 
from the spacetime metric  (\ref{metric}) and can be written in the 
usual Robertson-Walker form,
\beq
ds_{induced}^2=-dt^2+a_0^2 d{\bf x}^2,
\eeq
where the time coordinate has been redefined such that $n(t,0)=1$, 
and $a_0(t)\equiv a(t,y=0)$. $t$ is thus the proper time associated with the 
reference brane.

To get the induced geometry for the second brane, one must take into account
its displacement in the fifth dimension, and one finds from (\ref{metric})
the induced line element  
\beq
ds_{induced}^2=-\left[n^2(t,\y(t))-\dot\y^2\right]dt^2+a^2(t,\y(t))d{\bf x}^2,
\eeq
where a dot will always stand for a (possibly partial) derivative with 
respect to $t$.
This can be rewritten in the form 
\beq
ds_{induced}^2=-d\tau^2+a_2^2(\tau)d{\bf x}^2,
\eeq
where $\tau$ is the second brane cosmic time (i.e. the proper time for a 
comoving observer) which is related to the reference brane cosmic time $t$
by 
\beq \label{time}
d\tau=n(t,\y(t))\sqrt{1-{ \dot\y^2\over n^2(t,\y(t))}}\ dt\equiv
n_2 \gamma^{-1}\, dt.
\eeq
And the scale factor in the second brane is simply
\beq
a_2(t)=a(t,\y(t)),
\eeq
or $a_2(\tau)$ if one expresses $t$ in terms of $\tau$.

It is also useful to introduce  the expansion rate for the second brane,
which can be defined as 
\beq
H_2(t)\equiv {da_2/ dt\over a_2}= \left({\dot a\over a}+{a'\over a}\dot\y
\right)_2. \label{H2}
\eeq
However, $H_2$ does not coincide with the standard definition of the 
 Hubble parameter 
for a cosmological observer in the second brane because it is defined 
with respect to the time $t$, which is the cosmic time in the reference 
brane but not in the second brane. 
The traditional  Hubble parameter for the second brane corresponds to the 
definition 
\beq
{\cal H}_2(\tau)\equiv {da_2/ d\tau\over  a_2}={\gamma\over n_2}H_2
\label{calH2}
\eeq

Finally, one can introduce the unit velocity vector corresponding to the 
second brane, whose components read 
\beq
u^A=\{ {dt\over d\tau}, 0, 0, 0, {dy\over d\tau}\}=n_2^{-1} \gamma 
\{1, 0, 0, 0, {d\y\over dt}\}.
\label{velocity}
\eeq

At this point, we can easily infer the cosmological evolution 
in the second brane from  the knowledge of the cosmological evolution 
in the first brane and the position of the second brane with respect to the
first. What will be studied next is  the connection with the matter content 
of the second brane, which has been ignored up to this point. 

%***************************************************************************
\section{Junction conditions} \label{section4}
%***************************************************************************
The matter content of the  
branes is directly related to the jump of the extrinsic curvature tensor 
across the brane. 
This relation has been established explicitly in the case of a brane at 
rest with respect to the coordinate system (\ref{metric}) in our 
previous works 
\cite{bdl99,bdel99}. We generalize our result to include the case of a brane 
moving with respect to the coordinate system. 

We will now consider explicitly the second brane, but all the following 
expressions apply as well to the reference brane by taking $\y=0$.
The extrinsic curvature tensor on the  brane 
is  defined by the expression  
\beq
K_{AB}=h_A^C\nabla_C n_B, \label{extcurv}
\eeq
where 
$n^A$ is a unit vector field normal to the brane worlsheet and  
\beq
h_{AB}= g_{AB}- n_A n_B
\eeq
is the induced metric on the  brane.

It is easy to compute the components of the unit normal vector  from the 
components of the brane velocity (\ref{velocity}), 
by using the fact that $n^A$ must satisfy
the two following conditions:
\beq
g_{AB} n^An^B=1, \qquad g_{AB}n^A u^B=0,
\label{conds_normal}
\eeq
and  we get 
\beq
n^A=\{ {\dot \y\over n^2\sqrt {1-{\dot \y^2\over n^2}}},0,0,0, 
{1\over \sqrt {1-{\dot \y^2\over n^2}}} \}. 
\label{normal}
\eeq
Then, a straightforward calculation consisting in the  substitution of the 
components (\ref{normal}) into the definition (\ref{extcurv}) yields
the following (non zero) components for the extrinsic curvature tensor:
\begin{eqnarray} \label{K00}
K_0^0&=&{\ddot \y+nn'-2{n'\over n}\dot \y^2-{\dot n\over n}\dot \y\over
n^2\left(1- {\dot \y^2\over n^2}\right) ^{5/2}}, \\
K_0^5&=& \dot\y K_0^0, \qquad K_5^0=-K_0^5/n^2,\label{K05}\\
K^i_j&=&{1\over \sqrt {1-{\dot \y^2\over n^2}}}\left({a'\over a}
+{\dot a\dot \y\over a n^2}\right)\delta^i_j, \label{Kij}\\
K^5_5&=& - {\dot \y^2\left(
\ddot \y+nn'-2{n'\over n}\dot \y^2-{\dot n\over n}\dot \y\right) \over
n^4\left(1- {\dot \y^2\over n^2}\right) ^{5/2}}, \label{K55}
\end{eqnarray}
where all coefficients take their value on the  brane, i.e. at
$t$ and $y=\y(t)$.

Let us now introduce the energy-momentum tensor on the second brane, which, 
because of the symmetries of the setup, is necessarily of the perfect fluid
form 
\beq
S^{(2)}_{AB}=(\ro+\P)u_A u_B+\P h_{AB}.
\eeq
We also define $\hat S_{AB}$ as
\beq
\hat S_{AB} \equiv S_{AB}-{1\over 3}Sh_{AB}.
\eeq
The junction conditions \cite{Israel66} for a hypersurface in a
five-dimensional world read \cite{bdl99}
\beq
\left[ K_{\mu\nu}\right]=-\kappa^2\hat S_{\mu\nu},
\label{israel} 
\eeq
where the brackets stand for the jump across the brane,
and $K_{\mu\nu} = e_\mu^A e_\nu^B K_{AB}$ (and a similar expression for $\hat
S_{\mu\nu}$) where the four vectors $e^A_\mu$ ($\mu=0,1,2,3$) form a basis of
the vector space  tangent to the brane worldvolume. For the second brane, this
gives explicitly \beq
\left[ K_{\mu\nu}\right]_2=K_{\mu\nu}(t, \y(t)^+)-K_{\mu\nu}(t,\y(t)^-)
=-2K_{\mu\nu} (t,\y(t)),
\eeq
where the last equality follows from the identification of the point 
$y=\y(t)$ with the point $y=-\y(t)$ and from the $Z_2$ symmetry across
the second brane.
Note that, whereas all expressions (\ref{normal}-\ref{israel}) 
apply also to the reference brane (by taking $\y=0$), 
 the last expression differs by a sign for the two branes: one would find
$ \left[ K_{\mu\nu}\right]_0=2K_{\mu\nu} (t,y=0)$ for the reference brane.

It is easy to compute the components of $\hat S_A^B$, using the velocity 
components (\ref{velocity}) and one finds
\begin{eqnarray}
\hat{S}^0_0 &=& -\frac{ 2 \ro+3\P}{3}\gamma^2 \\
\hat{S}^0_5 &=& \frac{2 \ro + 3 \P}{3}\gamma^2{\dot{\y}\over n^2},
\quad \hat{S}^5_0=-\frac{2 \ro + 3 \P}{3}\gamma^2 \dot{\y}\\
\hat{S}^i_j &=& \frac{1}{3} \ro \delta^i_j\\
\hat{S}_5^5 &=& \frac{2 \ro + 3 \P}{3}\gamma^2{\dot{\y}^2\over n^2}.
\end{eqnarray}
Equating this  with 
the extrinsic curvature tensor, according to (\ref{israel}), 
all equations reduce to  only two 
 equations, which read 
\begin{eqnarray}
\frac{\ddot \y}{n^2} + \frac{n^\prime}{n}
\left(1- 2 \frac{\dot{\y}^2}{n^2}\right) -
\frac{\dot{n}}{n}\frac{\dot \y}{n^2} &=& -{\kappa^2\over 6} \left(2\ro + 3\P
\right)\left(1-\frac{\dot \y^2}{n^2} \right)^{3/2} 
\label{eom1}\\
\frac{a^\prime}{a}+ \frac{\dot a}{a}\frac{\dot \y}{n^2} &=&  {\kappa^2\over 6}
\ro\left(1-\frac{\dot {\y}^2}{n^2}\right)^{1/2},
\label{eom2}
\end{eqnarray}
where, once more, all the metric coefficients take their value on the brane.
The above system describes the full {\it non-linear dynamics of the 
homogeneous cosmological radion}. The non-linearity is manifest in the 
terms quadratic in $\dot \y$ but is also hidden in the fact that all 
the metric coefficients depend on $\y$.

It is interesting to note that the left hand side of 
 equation (\ref{eom1}) is simply the (covariant) acceleration in the 
fifth direction of a particle with the velocity (\ref{velocity}). This can 
be understood by noticing that $u^A K_{AB}u^B=-n_Bu^C \nabla_C u^B$,  
as a consequence of  (\ref{extcurv}) and (\ref{conds_normal}). 
$u^A K_{AB}u^B$, which is also $-K^0_0$, therefore vanishes for a geodesic 
motion. 
The right hand side of equation (\ref{eom1}) will thus induce
a deviation  from geodesic motion for a ``comoving''
observer on the second brane (at fixed coordinate $x^i$).

The second order equation (\ref{eom1}) can in fact be 
derived from the first order equation (\ref{eom2}). Indeed, if one 
takes the time derivative of (\ref{eom2}) and adds it to $H_2$ times 
(\ref{eom2}), one gets, after an overal multiplication by $\gamma^{-2}$, the
expression
\begin{eqnarray}
&& H_2\left[\frac{\ddot \y}{n^2} + \frac{n^\prime}{n}
\left(1- 2 \frac{\dot{\y}^2}{n^2}\right) -
\frac{\dot{n}}{n}\frac{\dot \y}{n^2}
\right]  \cr
&&-{1\over 3} G_{05}\left(1-{\dot\y^4\over n^4}\right)
-{1\over 3}\left(n^{-2}G_{00}+G_{55}\right)\left(1-{\dot\y^2\over n^2}\right)
\dot \y \cr
 &=& -{\kappa^2\over 6}H_2 \left(2\ro + 3\P
\right)\left(1-\frac{\dot \y^2}{n^2} \right)^{3/2},
\label{derivee}
\end{eqnarray}
where we have used 
\beq
\dot\ro=-3 H_2\left(\ro+\P\right),
\label{conservation2}
\eeq
 which expresses 
the energy conservation on the second brane (assuming there
is no inflow/outflow of energy from/to the bulk). 
A bulk energy-momentum tensor, satisfying the three-dimensional symmetries 
of homogeneity and isotropy, can be written in the form
\beq
T_{AB}= \rho_B v_A v_B+ P_T \left( h^{(2)}_{AB} + v_A v_B \right)  
+ P_B \  h^{(3)}_{AB},
\eeq
where $v^A$ is a unit time-like vector (whose ordinary spatial components 
are zero) describing the fluid velocity, $P_T$ represents the 
transverse pressure (along the fifth dimension) and $P_B$ represents
the  pressure in the three ordinary dimensions; $h^{(2)}_{AB}$ 
is the  projection tensor on the two-dimensional 
$(t,y)$ subspace and $h^{(3)}_{AB}$ is the 
projection tensor on the three-dimensional ordinary space. Because of 
the $Z_2$ symmetry, the bulk fluid cannot go ``across'' the branes, and we
must thus impose the further condition on the bulk fluid that the vector 
field $v^A$ coincides, on the branes,  with the brane velocity $u^A$
\footnote{ Simultaneously  $h^{(2)}_{AB} + v_A v_B = n_A n_B$.} . 
It is then not difficult to check explicitly that the second and third 
terms on the left hand side of (\ref{derivee}), expressed in terms of 
$T_{AB}$ by use of  Einstein's equations  (\ref{einstein}), just cancel.
Equation (\ref{eom1}) is thus derived, provided $H_2$ is not zero.

Equation (\ref{eom2}) can be solved algebraically to express 
$\dot\y$ in terms of the other quantities. One finds
\begin{eqnarray}
\dot{\y}&=& n\left(-\frac{a^\prime \dot{a}}{a^2n} \pm \frac{\kappa^2
\rho_2}{6}\sqrt{\frac{\dot{a}^2}{a^2 n^2} - \frac{a^{\prime 2}}{a^2}+
\frac{\kappa^4 \rho_2^2}{36}} \right) \left(\frac{\dot{a}^2}{n^2 a^2} +
\frac{\kappa^4 \rho_2^2}{36} \right)^{-1}.
\label{eom3}
\end{eqnarray}
If $a'/a$ and $\ro$ are of the opposite sign, then (\ref{eom2}) has 
one solution  if $|\dot a/(an)|>|a'/a|$, no solution otherwise.
If $a'/a$ and $\ro$ are of the same sign, then (\ref{eom2}) has 
two solutions when $|\dot a/(an)|< |a'/a|$, and only one solution otherwise. 
In both cases, when there is a unique solution, it corresponds 
to the root in (\ref{eom3}) with the sign being the same as that of 
$\dot a/a$.

To conclude this section, let us mention that the equations of motion 
for the radion can be rewritten in  terms of the cosmic time on the 
second brane, $\tau$, instead of the cosmic time on the reference brane 
$t$~: 
\begin{eqnarray}
\y_{,\tau\tau}+{n'\over n}\left(1+\y_{,\tau}^2\right) 
&=&
-{\kappa^2\over 6} \left(2\ro + 3\P
\right)\sqrt{1+\y_{,\tau}^2}, \label{33}\\
{a'\over a}+{\dot a\over an}{\y_{,\tau}\over \sqrt{1+\y_{,\tau}^2}}
&=&
{\kappa^2\over 6} \ro\left(1+\y_{,\tau}^2\right)^{-1/2}. \label{34}
\end{eqnarray}
The second equation has been obtained by \cite{ps00} in the particular 
case where the reference brane is Minkowski (hence $\dot a/a=0$ and 
$\rho_2$ is a constant.

%**********************************************************************
\section{Radion dynamics in an $AdS_5$ bulk spacetime}
%**********************************************************************
In order to explore the dynamical evolution of the radion, we now need 
to be more specific and to resort to explicit expressions for the spacetime
metric (\ref{metric}). We will thus choose to work with an 
$AdS_5$ bulk spacetime,
for which it is possible to recover (approximately) 
\cite{bdel99}, at late times, standard
cosmology, and which corresponds to the cosmological, i.e. non static, 
generalization of the Randall-Sundrum model \cite{rs99b}.

Keeping aside  the two branes, we assume 
 that the bulk spacetime is empty, except for a 
 negative cosmological  constant $\Lambda$.
It is useful to define a mass scale $\mu$ associated with $\Lambda$, according
to the expression
\beq
 \mu=\sqrt{-\Lambda/6},
\eeq
($\mu^{-1}$ is the $AdS_5$ curvature radius) 
and an energy density scale  $\sigma$, defined 
by
\beq
\sigma=6{\mu\over \kappa^2}.
\eeq
Note that, in the Randall-Sundrum model \cite{rs99b}, 
the energy density in the brane is supposed to be exactly this 
 $\sigma$. In the present work, we consider any value for the energy densities
in the two branes. 

 We will also 
assume that the five-dimensional Weyl tensor is zero. Our global spacetime is 
therefore made of two pieces of $AdS_5$ separated by the two branes.
Assuming the usual $Z_2$ symmetry with respect to each brane, the two
$AdS_5$  pieces are then simply mirror symmetric.

In \cite{bdel99}, we have solved explicitly the five-dimensional 
Einstein's equations 
\beq
G_{AB}-\Lambda g_{AB}=\kappa^2 T_{AB},
\eeq
in the case where the only matter in spacetime is the one confined in a brane.
We obtained the following form for the component of (\ref{metric}):
\begin{eqnarray}
a(t,y) &=& a_0(t)\left(\cosh{\mu y} -\eta_0 \sinh{\mu |y|}\right) \label{a}\\
n(t,y) &=& \cosh{\mu y} -\tilde\eta_0 \sinh{\mu |y|}, \label{n}
\end{eqnarray}
where 
 $\eta_0$ and $\tilde\eta_0$ are two functions of time,
related to the energy density in the reference brane by
\beq
\eta_0(t)={\rho_0(t)\over\sigma},\qquad
\tilde\eta_0(t)=\eta_0+{\dot\eta_0 \over H_0},
\label{lambda}
\eeq
where 
$H_0\equiv\dot a_0/a_0$ is the Hubble parameter.
For future use we also define $\eta_2$ as
\beq
\eta_2(t)={\rho_2(t)\over\sigma},
\eeq
where $\rho_2(t)$ is the energy density of the second brane.
It must be stressed that, whereas the functional form of the metric
components (\ref{a},\ref{n}) is obtained by solving the Einstein equations in the 
bulk, the relationship between the coefficients and the reference brane 
matter given in (\ref{lambda}) comes from the junction conditions 
\cite{bdl99} across the reference brane.

Finally, the evolution of the scale factor $a_0(t)$ and of the energy 
density $\rho_0(t)$ is determined by solving the coupled system consisting 
of the generalized Friedmann equation
\beq
H_0^2={ \Lambda \over 6}+{ \kappa^4 \over 36}\rho_0^2
- {k \over a_0^2} =  \mu^2\left(\eta_0^2-1\right) -{k \over a_0^2}
 \label{fried}
\eeq
where $k=-1,0,1$ according to the three-dimensional spatial curvature
(the term ${\cal C}/a_0^4$  does not appear because we have assumed 
a vanishing Weyl tensor) and the usual energy conservation law
\beq
\dot\rho_0=-3H_0\left(\rho_0+P_0\right). \label{conservation}
\eeq
%***[In the particular case where $\dot{\eta}_0 =0$, one can easily verify from
%(\ref{fried}) and (\ref{conservation}) that the induced geometry in the
%reference brane is $dS_4$, Minkowkian or $AdS_4$ according to the position of
%$\eta_0^2$ with respect to $1$.]***

A useful identity is an  integral of Einstein's equations
\beq
\left(\frac{\dot a}{na}\right)^2 = \frac{1}{6}\kappa^2 \rho_B +
\left(\frac{a^\prime}{a}\right)^2 - \frac{k}{a^2} 
\eeq
which was obtained in \cite{bdel99}. The generalized Friedmann equation 
(\ref{fried}) on the reference brane is simply the application of this 
identity (with the junction conditions) in $y=0$.

 The fully non-linear evolution of the radion, coupled to the evolution 
of the matter on  the two branes, is thus given by
 the first-order system of differential equations consisting of 
(\ref{fried}), (\ref{conservation}), (\ref{conservation2}) and 
(\ref{eom3}), which can be rewritten in the simpler form
\begin{eqnarray}
\dot{\y}&=& n\left(-\frac{a^\prime \dot{a}}{a^2n} \pm \frac{\kappa^2
\rho_2}{6}|{\cal H}_2|\right) \left(\frac{\dot{a}^2}{n^2 a^2} +
\frac{\kappa^4 \rho_2^2}{36} \right)^{-1},
\end{eqnarray}
where ${\cal H}_2$ is the Hubble parameter of the second brane defined in 
(\ref{calH2}).

One can also check, starting from the expression (\ref{H2}) in terms 
of the metric component (\ref{a})
 and using (\ref{eom2}) and (\ref{fried}), 
all terms in $\dot\y$  cancel
and  one recovers the 
generalized Friedmann equation on  the second brane, that is
 \beq
{\cal H}_2^2={ \Lambda \over 6}+{ \kappa^4 \over 36}\ro^2
- {k \over a_2^2} = \mu^2\left(\eta_2^2-1\right) -{k \over a_2^2}
 \label{fried2}.
\eeq
It has to be so  because the 
generalized Friedmann equations comes from a purely local analysis, and the 
r\^ole of the two branes can be exchanged.

%**********************************************************************
\section{Radion fluctuations}
%**********************************************************************
In this section, we  study perturbatively 
 the dynamics of the radion fluctuations. We first define 
 the background solution as the
case where the radion is frozen, i.e. does not depend on time.
This means that  the second brane is 
at rest with respect to the first brane, i.e. 
\beq
y_2(t)=\bar\y, 
\eeq
where $\bar \y$ is a constant.
 One can then 
find explicit solutions of the Einstein's equations for the whole two-brane
system. However, to ensure the continuity of the metric along the compact 
fifth dimension, one finds that the matter content in the second brane 
is necessarily related to the matter content of the first brane. 
This result was already noticed for a vanishing bulk cosmological constant 
\cite{bdl99}. We generalize it here to the case of a non-zero bulk cosmological 
constant.

If $\y$ is constant, then the system (\ref{eom1}-\ref{eom2}) simplifies 
considerably to yield
\beq
\left(\frac{n^\prime}{n}\right)_2= -{\kappa^2\over 6} 
\left(2\bar \ro + 3\bar\P
\right)
\eeq
and
\beq
\left(\frac{a^\prime}{a}\right)_2={\kappa^2\over 6}\bar\ro.
\eeq
Substituting the explicit expressions (\ref{a}) and (\ref{n}) for 
$a$ and $n$, which depend on the energy content on the first brane, one 
can express  the energy density and pressure of the second brane as functions
of the energy density and pressure on the first brane and of the position 
$\y$. 
One finds
\beq \label{barrho}
\kappa^2\bar\ro(t)=6\mu {\sinh\mu\bar\y-\eta_0\cosh\mu\bar\y\over
\cosh\mu\bar\y-\eta_0\sinh\mu\bar\y},
\eeq
and
\beq \label{barp}
\kappa^2\bar\P(t)= -2\mu{
\sinh\mu\bar\y-\tilde\eta_0\cosh\mu\bar\y\over
\cosh\mu\bar\y-\tilde\eta_0\sinh\mu\bar\y}-{2\over 3}\kappa^2\bar\ro.
\eeq
Note that if one considers an equation of state of the form $P_0=w_0\rho_0$
with $w_0$ constant for the matter on the reference brane (which  includes
the standard cases of radiation, non-relativistic matter and a cosmological 
constant), the equation of state of the matter on the second brane, 
which is completely determined here as a function of the relative distance 
between the two branes, will correspond  to an equation of state 
$\bar{P}_2=w_2\bar{\rho}_2$ where, in general, $w_2$ is time dependent. Indeed
$\eta_0$ and
\beq
\tilde\eta_0= -(2+3w_0)\eta_0
\eeq
are time dependent in general. 

One exception is when the reference brane 
contains only a cosmological constant, i.e. $w_0=-1$, then one also gets 
$w_2=-1$ on the second brane.
There is in this case an interesting correspondence between the
cosmological constants on the two branes. Introducing the effective
cosmological constant from the brane point of view as
\begin{equation}
\lambda = {\Lambda \over 6} + {\kappa^4 \over 36} \rho = \mu^2 (\eta^2 -
1), \label{efflambda}
\end{equation}
we have
\begin{equation}
\lambda_2 = {\lambda_0 \over \left( \cosh\mu\bar\y-\eta_0\sinh\mu\bar\y 
\right)^2}.
\end{equation}
Thus, the two branes are simultaneously $dS_4$ ($\lambda>0$, 
$|\eta| >1$), $M_4$ ($\lambda=0$, $|\eta| =1$) or $AdS_4$ ($\lambda<0$, 
$|\eta| <1$) \cite{kl99}.

 Another interesting case is when the
reference brane undergoes {\it ordinary cosmology}, which occurs when 
$\eta_0$ is very small. One can then neglect $\eta_0$ as well as $\tilde\eta_0$
in the  expressions (\ref{barrho}-\ref{barp}) 
and the matter on the second brane reduces effectively
to a cosmological constant.

In order to obtain the equation of motion for the radion fluctuations, 
we linearize the exact equations of motion (\ref{eom1}-\ref{eom2}) about
the equilibrium configuration we have just defined. 
The linearized system  reads
\begin{eqnarray}
\frac{\ddot \dy}{n^2} + \delta \left(\frac{n^\prime}{n}\right) -
\frac{\dot{n}}{n}\frac{\dot \dy}{n^2} &=& -{\kappa^2\over 6} 
\left(2\delta\ro + 3\delta\P\right)
\label{deom1}\\
\delta\left(\frac{a^\prime}{a}\right)+ \frac{\dot a}{a}\frac{\dot \dy}{n^2} 
&=&  {\kappa^2\over 6}
\delta\ro,
\label{deom2}
\end{eqnarray}
where we recall that all metric coefficients take their value on the 
second brane.
Furthermore, one can use the explicit expressions (\ref{a}-\ref{n}) to 
find
\begin{eqnarray}
\delta\left(\frac{n^\prime}{n}\right)&=& m_n^2\dy\equiv 
\left[\mu^2-{\kappa^4\over 36}\left(2\bar\ro+3\bar\P\right)^2\right]\dy \\
\delta\left(\frac{a^\prime}{a}\right)&=&
m_a^2\dy\equiv
 \left[\mu^2-{\kappa^4\over 36}\bar\ro^2\right]\dy,
\end{eqnarray}
where we have introduced two effective squared masses $m_n^2$ and $m_a^2$.
Note that both $m_n$ and $m_a$ vanish in the particular case where the  
Randall-Sundrum condition between the tension and the bulk cosmological 
constant  is satisfied on the second brane. 

If one adds (\ref{deom1}) with three times (\ref{deom2}), one obtains
the equation of motion 
\beq 
\frac{\ddot \dy}{n^2} +\left(3\frac{\dot a}{a}-
\frac{\dot{n}}{n}\right)\frac{\dot \dy}{n^2}+\left(m_n^2+3m_a^2\right)\dy=
-{\kappa^2\over 6} \delta T_2, 
\label{tracefluct}
\eeq
where $\delta T_2\equiv -\d\ro+3\d\P$ is the linear perturbation of the 
trace of the matter energy-momentum tensor. 
One thus recognizes the equation of motion for a scalar field coupled to 
the (perturbed) trace of the matter  energy-momentum tensor. The friction 
coefficient is the usual one 
(rewriting in terms of $\tau$
would give the natural $3{\cal H}_2$ friction coefficient, as can be
seen when extracting this equation directly from (\ref{33}), (\ref{34})), 
but the mass term
is an effective mass that depends both on the bulk cosmological 
constant and on the background matter content of the second brane. 

This 
effective mass is time-dependent except in the case where the second brane 
matter has for equation of state $\P=-\ro$. The two squared masses are then 
identical and equal to
\beq \label{masse}
m_a^2=m_n^2=-\lambda_2,
\eeq
where $\lambda_2$ is the effective cosmological constant from the brane 
point of view defined in (\ref{efflambda}). 
One thus recovers exactly previous results \cite{cgr99,gs00,kr00}.
An immediate consequence of this is that when the first brane is $dS_4$,
the small perturbations around the equilibrium position of the second brane
(then also $dS_4$) are unstable, whereas when the first brane is $AdS_4$ (the
second brane is then also $AdS_4$) they are stable. We will see in section
\ref{section7} that this perturbative analysis matches with the exact non
linear evolution. 

The evolution of the radion fluctuations is  entangled with the 
matter fluctuations. It is however possible when the matter fluctuations are 
 of the adiabatic type, i.e. 
\beq
\delta\P=c_{s2}^2\delta\ro,
\eeq
to derive an equation for only the radion fluctuations by considering the 
appropriate linear combination of (\ref{deom1}) and (\ref{deom2}), 
\beq
\frac{\ddot \dy}{n^2}+\left[\left(2+3c_{s2}^2\right){\dot a\over an^2}
-{\dot n\over n^3}\right]\dot\dy+\left[m_n^2+\left(2+3c_{s2}^2\right)m_a^2
\right]\dy=0. 
\label{flucteom1}
\eeq
The evolution of the matter fluctuations is then related to the radion 
fluctuation by (\ref{deom2}), i.e.
\beq
{\kappa^2\over 6}\d\ro=\frac{\dot a}{a}\frac{\dot \dy}{n^2}+
m_a^2\dy. 
\label{flucteom2}
\eeq

Assuming  an equation of state $\bar \P=w_2\bar\ro$ 
(and adiabatic fluctuations, i.e. 
$c_{s2}^2=w_2$), the equation of motion (\ref{flucteom1}) takes the form,
after substitution of $m_a^2$ and $m_n^2$ by their explicit expressions,
\beq
\dy_{,\tau\tau}+(2+3w_2){\cal H}_2
\dy_{,\tau}+3(1+w_2) \left(\mu^2- {\kappa^4\over
36}\bar{\ro}^2(2+3w_2)\right)\dy=0, \eeq
where we have used here the second brane cosmic time $\tau$. 
One can solve this equation for de Sitter ($w_2=-1$) on the two branes and
one gets 
\beq 
\dy=\dy_i e^{{\cal H}_2(\tau-\tau_i)}-{\kappa^2\over 6 {\cal
H}_2^2}\delta{\rho}_{2i} \left(1- e^{{\cal H}_2(\tau-\tau_i)}\right),
\eeq
where $\dy_i$ is the initial radion fluctuation and $\delta{\rho}_{2i}$
is the initial density fluctuation (at time $\tau_i$). For de Sitter brane
cosmology, we recover the instability mentioned above.

Note that in the case of the radion fluctuations in the strict Randall-Sundrum 
model, studied in \cite{cgr99}, the friction coefficient in the above 
equation vanishes because the background is static, and both the effective
squared masses $m_a^2$ and $m_n^2$ also vanish because of the RS condition 
relating the bulk cosmological constant and the brane tensions. One thus 
ends up with the equation of motion for a free scalar field.

%%%%%%%%%%%%%%%%%%%%%%%%%%%%%%%%%%%%%%%%%%%%%%%%%%%%%%%%%%%%%%%%%%%%%%%%%%%
\section{Non linear evolution of the radion} \label{section7}
%%%%%%%%%%%%%%%%%%%%%%%%%%%%%%%%%%%%%%%%%%%%%%%%%%%%%%%%%%%%%%%%%%%%%%%%%%%

In this section, we explore the non-perturbative aspects of the radion 
evolution in some simple cases, leaving a more systematic analysis for 
future work.
We are going to consider the cases where the matter on the reference brane
 behaves like a cosmological constant. In these simple cases, the
reference brane matter does not evolve, i.e.
$\dot\eta_0=0$, and the metric components are separable
and can be written in the form
\beq
n(t,y)=n(y), \qquad a(t,y)=a_0(t) n(y).
\eeq
The explicit form of $n(y)$ depends on the sign of $\eta_0-1$. If 
$\eta_0>1$, the brane is \dS; if $\eta_0<1$, the brane is \AdS; and 
the case $\eta_0=1$ corresponds to the Randall-Sundrum model \cite{rs99b}.
We will then further specialize to the cases where the brane matter of the
second also behaves as a cosmological constant, implying $\dot{\eta}_2=0$.

%%%%%%%%%%%%%%%%%%%%%%%%%%%%%%%%%%%%%%%%%%%%%%%%%%%%%%%%%%%%%%%%%%%%%%%%%%%%
\subsection{$dS_4$ brane}

For a $dS_4$  reference brane, the dependence of the  metric components
on the fifth coordinate is simply (see e.g. \cite{lmw00}) 
\beq
n(y)=\sqrt{\eta_0^2-1}\sinh\mu(y_{\rm h}-|y|),
\eeq
where 
\beq
y_{\rm h} ={1\over \mu}\arg \coth \eta_0
\eeq
Here we suppose that the second brane is always located at $\y < y_{\rm
h}$. Otherwise, there would be a horizon at $\y = y_{\rm h}$.

Equation (\ref{eom3}) can then be rewritten in terms of dimensionless 
quantities in the form
\beq
\label{dsx}
x_{,s}= - {\cal K}_0^2 \sinh x \frac{h_0 \cosh x \pm
\eta_2 \sinh x \sqrt{\eta_2^2 {\cal K}_0^2 \sinh^2 x
- {\cal K}_0^2 \cosh^2 x +h_0^2}}{h_0^2 + \eta_2^2 {\cal K}_0^2
\sinh^2 x}
\eeq
where we have introduced the dimensionless distance of the
second brane to the horizon $x$ and  the dimensionless time $s$
\beq \label{dimless_time}
x\equiv \mu(y_{\rm h}-\y),\quad s=\mu t.
\eeq
The dimensionless Hubble parameter $h_0$ is defined by
\beq
h_0 \equiv \frac{1}{a_0}\frac{da_0}{ds} = \frac{H_0}{\mu},
\eeq
and ${\cal K}_0$ is defined by
\beq \label{K}
{\cal K}_0 = \sqrt{|\eta_0^2-1|}
\eeq
In general equation (\ref{dsx}) is non separable
since both $\eta_2$ and $h_0$ are function of time. It gets
however simplified
when one assumes that the energy momentum tensor of the second brane is that of
a cosmological constant (implying $\dot{\eta}_2=0$),  and if one assumes
moreover that $k=0$ (in order to account for a slicing of $dS_4$ by flat
3-spaces) so that $h_0= {\cal K}_0^2$  (as can be seen
from (\ref{fried})). In this case equation
(\ref{dsx}) reads
\beq \label{motionDS} x_{,s}=-\sqrt{\eta_0^2-1}\sinh x{\cosh
x\pm \eta_2\sqrt{\eta_2^2-1} \sinh^2 x\over 1+\eta_2^2\sinh^2 x},
\label{eomdS}
\eeq
it becomes separable and can be solved analytically. 
Since the matter energy densities are frozen on both branes,
the evolution of the radion is fully embodied in this unique first order
equation. The evolution of the radion is thus completely determined by the
knowledge of the initial position of the second brane, which we call $\y_i$. If
the initial position $\y_i$ is located at the equilibrium position \beq
\bar{x}=\mu(y_{\rm h}-\bar{\y})=-\arg \coth \eta_2,
\eeq
then the brane does not move. However, if the brane is initially away 
from the equilibrium position, it moves. According to the perturbative 
analysis, a brane slightly off the equilibrium position  moves further 
away since we found that the radion is unstable. 

\begin{figure}
\begin{center}
\epsfxsize=4.8 in \epsfbox{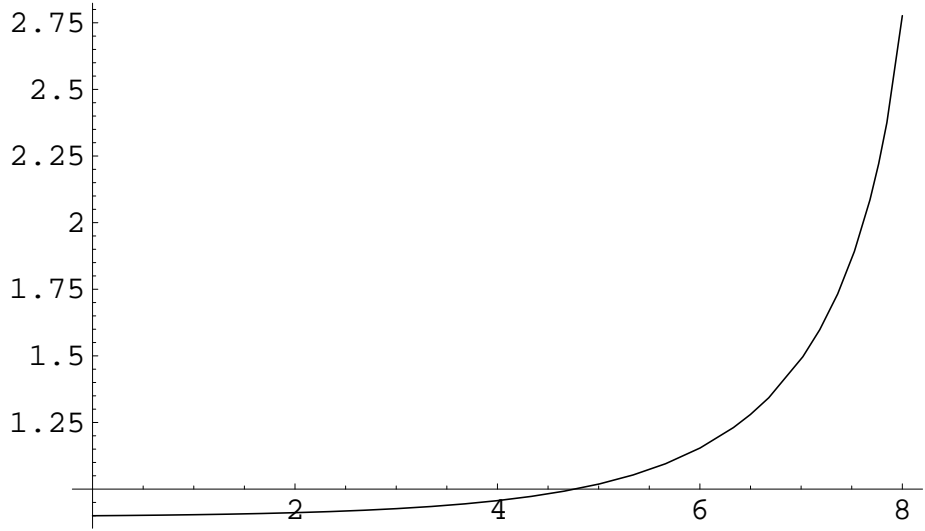}
\end{center}
\caption{Evolution with time of the distance $x$ from the second brane to the
``horizon''  in the case of $dS_4$ for an initial position $\y_i < \bar
\y$. 
 \label{figure1}}
\end{figure}

\begin{figure}
\begin{center}
\epsfxsize=4.8 in \epsfbox{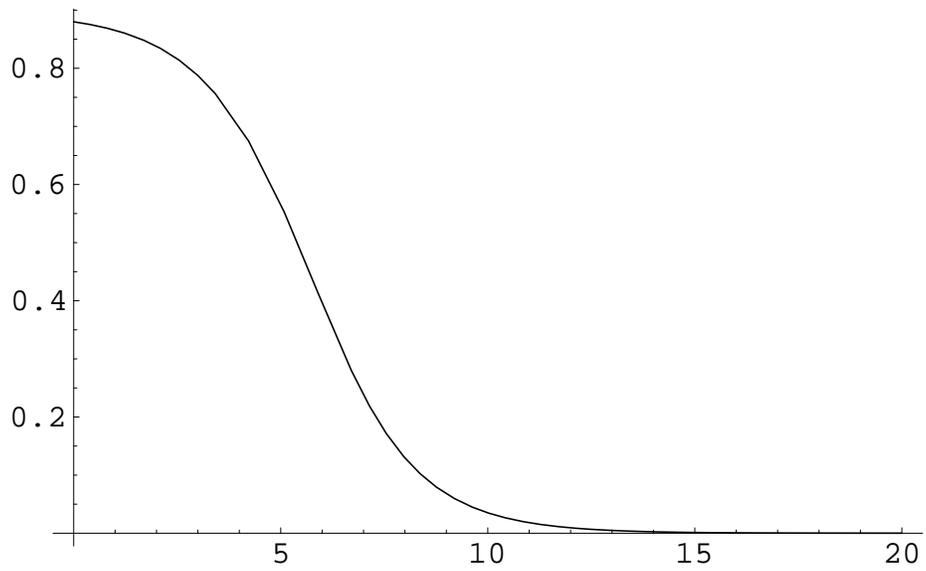}
\end{center}
\caption{Evolution with time of the distance $x$ from the second brane to the
``horizon''  in the case of $dS_4$ for an initial position $\y_i > \bar
\y$. 
\label{figure2}}
\end{figure}

Here, we can determine the non-linear evolution.
We have solved numerically the evolution of the radion in the cases where 
the brane is initially on the left ($\y_i<\bar{\y}$) and on the right
($\y_i>\bar{\y}$) of the equilibrium position.
We observe the expected unstable behaviour. In the first case, 
illustrated in Fig. \ref{figure1},  the brane
moves towards the reference brane until it collides with it (numerically 
it goes across it). In the second case, illustrated in Fig. \ref{figure2},
 the brane moves further away 
but its motion freezes asymptotically at the horizon, as seen in the
reference brane proper time. However, as can be seen from (\ref{time}) and
(\ref{motionDS}), this takes only a finite amount of the second brane 
proper time, as is typical for the crossing of a (Rindler or black
hole) horizon.

One may obtain a simple analytical expression which illustrates this 
behaviour in the case of a empty second brane~: $\eta_2 = 0$. The
solution is 
\begin{equation}
x = \arg \tanh  \left[ \tanh x_i \exp \left( - \sqrt{\eta_0^2 -1}
(s-s_i) \right) \right],
\end{equation}
where $x_i$ is the initial position at time $s_i$. This is also the solution 
of (\ref{eomdS}) for non zero $\eta_2$ when the second brane approaches 
the horizon (i.e. for $x$ small). 

%%%%%%%%%%%%%%%%%%%%%%%%%%%%%%%%%%%%%%%%%%%%%%%%%%%%%%%%%%%%%%%%%%%%%%%%%%%%

%%%%%%%%%%%%%%%%%%%%%%%%%%%%%%%%%%%%%%%%%%%%%%%%%%%%%%%%%%%%%%%%%%%%%%%%%%%%
%%%%%%%%%%%%%%%%%%%%%%%%%%%%%%%%%%%%%%%%%%%%%%%%%%%%%%%%%%%%%%%%%%%%%%%%%%%%

%%%%%%%%%%%%%%%%%%%%%%%%%%%%%%%%%%%%%%%%%%%%%%%%%%%%%%%%%%%%%%%%%%%%%%%%%%%%

\subsection{$AdS_4$ brane}
For an $AdS_4$ reference brane one finds for the metric component
\beq
n(y)=\sqrt{1-\eta_0^2}\cosh\mu(|y|-y_c),
\eeq
where
\beq
y_c ={1\over \mu}\arg \tanh \eta_0.
\eeq
In this case the warp factor $n$ has a minimum in $y=y_c$ enabling to have a
configuration with two static branes of positive tensions \cite{kl99}
(see also \cite{kr00,Kogan:2001vb}).
In any case, equation
(\ref{eom3}) reads \beq \label{Adsx}
x_{,s} ={\cal K}_0^2 \cosh x \frac{-h_0  \sinh x \pm
\eta_2 \cosh x \sqrt{\eta_2^2 {\cal K}_0^2 \cosh^2 x
- {\cal K}_0^2 \sinh^2 x +h_0^2}}{h_0^2 + \eta_2^2{\cal K}_0^2
\cosh^2 x},
\eeq
where $x$ is defined by
\beq
x \equiv \mu (\y-y_c),
\eeq
and $s$ and ${\cal K}_0$ are defined respectively as in 
(\ref{dimless_time}) and
(\ref{K}).
The scale factor is given straighforwardly by solving
(\ref{fried}); one finds $a_0=\sin ({\cal K}_0 s) / \mu{\cal K}_0$, so that
in this FLRW parametrization of $AdS_4$ there are 
coordinate singularities in ${\cal K}_0 s=0$ 
and ${\cal K}_0 s=\pi$. The 
dimensionless Hubble parameter is then
 $h_0 = {\cal K}_0 \mbox{cotan} ({\cal K}_0s)$.
The equation of motion (\ref{Adsx}) remains non separable  even in the simple case where
$\dot{\eta_2}=0$, with $\eta_2 \neq 0$.
We have solved numerically for the evolution of the radion in the cases where
the brane (with $\eta_2$ being constant and positive)
 is initially (at time $s_i$ such that 
${\cal K}_0 s_i < \pi/2$)
on the left
($x_i<\bar{x}$) and on the right ($x_i>\bar{x}$) of the equilibrium position
$\bar{x}$ defined by
\beq
\bar{x}=  \mu(\bar{\y}-y_c) = \arg \tanh \eta_2.
\eeq
For small enough $s_i$, the only acceptable solution is given by taking the 
$+$ sign in front of the square root in equation (\ref{Adsx}), as can be seen using 
the arguments given at the end of section  \ref{section4}. 

\begin{figure}
\begin{center}
\epsfxsize=4.8 in \epsfbox{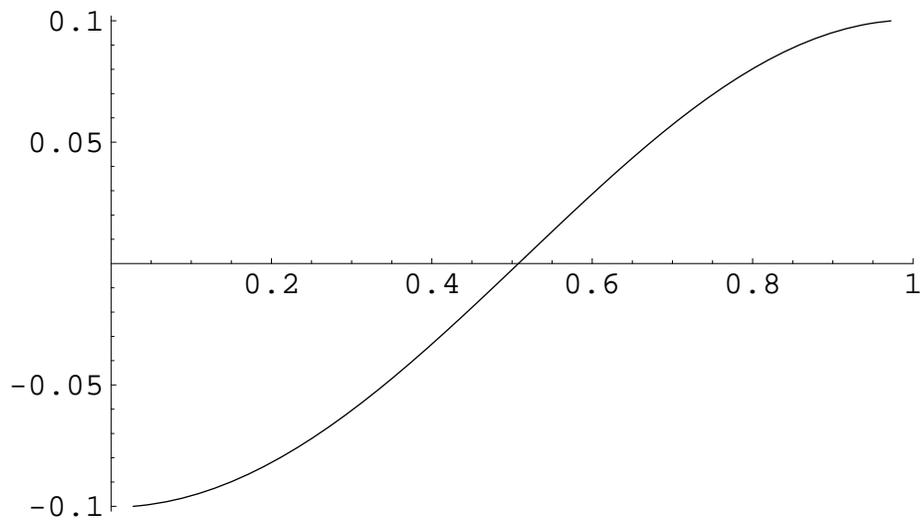}
\end{center}
\caption{ Evolution of the distance $x-\bar x$ between the brane
position and the equilibrium position
 as a function of ${\cal K}_0  s / \pi$,
in the case of $AdS_4$ for an initial
position $x_i< \bar x$.
\label{figure3}}
\end{figure}

\begin{figure}
\begin{center}
\epsfxsize=4.8 in \epsfbox{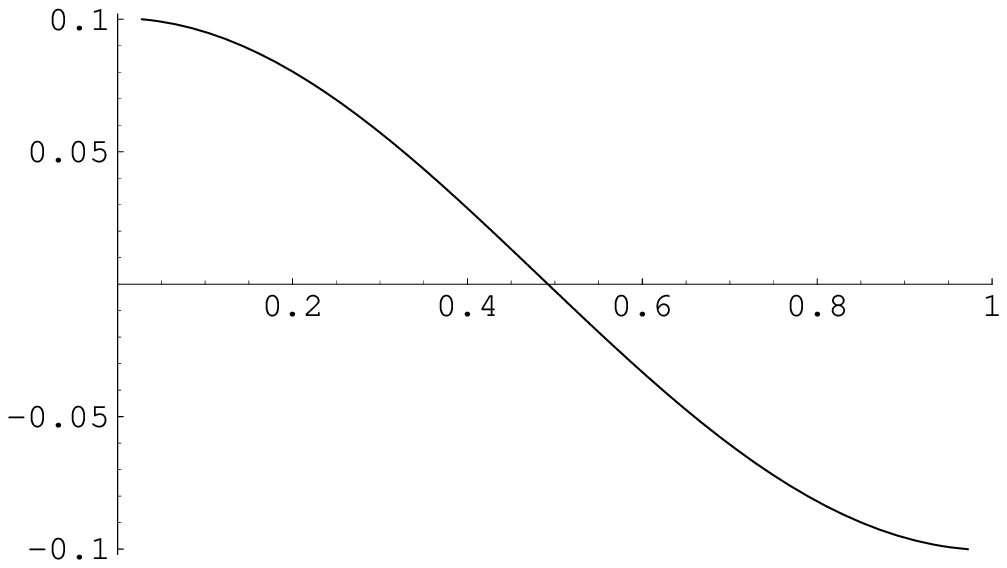}
\end{center}
\caption{ Evolution of the distance $x-\bar x$ between the brane
position and the equilibrium position
as a function of ${\cal K}_0  s / \pi$, 
in the case of $AdS_4$ for an initial
position $x_i > \bar x$.
\label{figure4}}
\end{figure}

In both cases ($x_i<\bar{x}$ and $x_i>\bar{x}$), illustrated in figures 
\ref{figure3} and \ref{figure4}, the
brane first moves back toward its equilibrium position, and
 then reaches a point (at time $s_c$ close to $ \pi/2{\cal K}_0$)
where the  square root\footnote{which is also the determinant of the quadratic 
equation obtained from equation  (\ref{eom2})} in equation 
(\ref{Adsx}) vanishes. The solution obtained from   (\ref{Adsx}) with
 the $+$ sign in front of the square root 
must then be matched (after time $s_c$) to a solution of (\ref{Adsx}) with 
 a $-$ sign is taken in 
front of the square root 
(the two solutions have the same tangent in $s_c$). 
Figures \ref{figure3} and \ref{figure4} are obtained with this matching.

Equation (\ref{Adsx}) becomes separable, and can be easily integrated, in the
limiting case where the second brane is empty i.e. $\eta_2 =0$. In this case,
one finds: \begin{equation}
x = \arg \tanh \left[ \tanh x_i { \cos \left({\cal K}_0 s \right) \over
\cos \left( {\cal K}_0 s_i \right)} \right].
\end{equation}

%%%%%%%%%%%%%%%%%%%%%%%%%%%%%%%%%%%%%%%%%%%%%%%%%%%%%%%%%%%%%%%%%%%%%%%%%%%%

%%%%%%%%%%%%%%%%%%%%%%%%%%%%%%%%%%%%%%%%%%%%%%%%%%%%%%%%%%%%%%%%%%%%%%%%%%%%
%%%%%%%%%%%%%%%%%%%%%%%%%%%%%%%%%%%%%%%%%%%%%%%%%%%%%%%%%%%%%%%%%%%%%%%%%%%%

%%%%%%%%%%%%%%%%%%%%%%%%%%%%%%%%%%%%%%%%%%%%%%%%%%%%%%%%%%%%%%%%%%%%%%%%%%%%
%%%%%%%%%%%%%%%%%%%%%%%%%%%%%%%%%%%%%%%%%%%%%%%%%%%%%%%%%%%%%%%%%%%%%%%%%%%%

\subsection{$M_4$ brane}
When the reference brane is Minkowskian
($\eta_0$ being 1, and $a=n=\exp (-\mu |y|)$)
, equation (\ref{eom3}) is given by
\beq \label{Mx}
x_{,s} = \pm \frac{e^{-x}}{\eta_2} \sqrt{\eta_2^2-1},
\eeq
where $x$ is given here by
\beq
x \equiv \mu \y
\eeq
and $s$ is defined as above. The second brane equilibrium condition is simply given by 
the Randall-Sundrum \cite{rs99a,rs99b} matching condition expressed here as $\eta_2=-1$.
For this particular $\eta_2$, one sees easily from equation (\ref{Mx}) 
that the second brane must  be at rest 
with respect to the first one, but that the equilibrium 
position $\bar x$ can be chosen arbitrarily. However a small departure from the equilibrium
 condition $\eta_2 = -1$ results in either a colliding of the two branes or 
a runaway behaviour.

\section*{Acknowledgments}
We thank Arthur Lue, who helped us to understand some subtleties associated 
with the numerical integration of equation (\ref{Adsx}). 
The work of C.D. is sponsored in part by NSF Award PHY 9803174,
 and by  David and Lucile Packard Foundation
Fellowship 99-1462.

\end{document}